\documentclass[aps,amsmath,twocolumn,amssymb,floatfix,showpacs,superscriptaddress,nofootinbib,longbibliography]{revtex4-1}
\usepackage{mathtools}
\usepackage{braket}
\usepackage[dvipsnames]{xcolor}
\usepackage{float}
\usepackage{subfigure}
\usepackage[colorlinks=true,linktoc=page,linkcolor=blue,citecolor=blue,urlcolor=blue]{hyperref}

\newcommand{\vect}[1]{\boldsymbol{\mathrm{#1}}}
\mathchardef\mhyphen="2D 

\newcommand{\ie}{{i.e.,\,\,}}
\newcommand{\eg}{{e.g.,~}}
\newcommand{\ua}{{\uparrow }}
\newcommand{\da}{{\downarrow }}

\newcommand\bea{\begin{eqnarray}}
\newcommand\eea{\end{eqnarray}}
\newcommand\beq{\begin{equation}}  
\newcommand\eeq{\end{equation}}

\newcommand{\non}{\nonumber}  
\usepackage[normalem]{ulem}
\definecolor{lime}{HTML}{A6CE39}
\usepackage{sidecap,tikz}
\DeclareRobustCommand{\orcidicon}{\hspace{-1.0mm}
	\begin{tikzpicture}
		\draw[lime, fill=lime] (0.0,0.0) 
		circle [radius=0.15] 
		node[white] {{\fontfamily{qag}\selectfont \tiny \,ID}};
		\draw[white, fill=white] (-0.0525,0.095) 
		circle [radius=0.007];
	\end{tikzpicture}
	\hspace{-3.0mm}
}
\foreach \x in {A, ..., Z}{\expandafter\xdef\csname orcid\x\endcsname{\noexpand\href{https://orcid.org/\csname orcidauthor\x\endcsname}
		{\noexpand\orcidicon}}
}

\AtBeginDocument{%
	\newwrite\bibnotes
	\def\bibnotesext{Notes.bib}
	\immediate\openout\bibnotes=\jobname\bibnotesext
	\immediate\write\bibnotes{@CONTROL{REVTEX41Control}}
	\immediate\write\bibnotes{@CONTROL{%
			apsrev41Control,author="08",editor="1",pages="1",title="1",year="1"}}
	\if@filesw
	\immediate\write\@auxout{\string\citation{apsrev41Control}}%
	\fi
}%
\begin{document}


\title{Topological characterization and stability of Floquet Majorana modes in Rashba nanowire}

\author{Debashish Mondal\orcidD{}}
\email{debashish.m@iopb.res.in}
\affiliation{Institute of Physics, Sachivalaya Marg, Bhubaneswar-751005, India}
\affiliation{Homi Bhabha National Institute, Training School Complex, Anushakti Nagar, Mumbai 400094, India}

\author{Arnob Kumar Ghosh\orcidA{}}
\email{arnob@iopb.res.in}
\affiliation{Institute of Physics, Sachivalaya Marg, Bhubaneswar-751005, India}
\affiliation{Homi Bhabha National Institute, Training School Complex, Anushakti Nagar, Mumbai 400094, India}

\author{Tanay Nag\orcidB{}}
\email{tanay.nag@physics.uu.se}
\affiliation{Department of Physics and Astronomy, Uppsala University, Box 516, 75120 Uppsala, Sweden}

\author{Arijit Saha\orcidC{}}
\email{arijit@iopb.res.in}
\affiliation{Institute of Physics, Sachivalaya Marg, Bhubaneswar-751005, India}
\affiliation{Homi Bhabha National Institute, Training School Complex, Anushakti Nagar, Mumbai 400094, India}

\begin{abstract}
We theoretically investigate a practically realizable Floquet topological superconductor model, based on a one-dimensional Rashba nanowire and proximity induced $s$-wave superconductivity in the presence of a Zeeman field. The driven system hosts regular $0$- and anomalous $\pi$-Majorana end modes~(MEMs). By tuning the chemical potential and the frequency of the drive, we illustrate the generation of multiple MEMs in our theoretical set up. We utilize the chiral symmetry operator to topologically characterize these MEMs via a dynamical winding number constructed out of the periodized evolution operator. Interestingly, the robustness of the $0$- and $\pi$-MEMs is established in the presence of on-site time-independent random disorder potential. We employ the twisted boundary condition to define the dynamical topological invariant for this translational-symmetry broken system. The interplay between the Floquet driving and the weak disorder can stabilize the MEMs giving rise to a quantized value of the dynamical winding number for a finite range of drive parameters. This observation might be experimentally helpful in scrutinizing the topological nature of the Floquet MEMs. We showcase another driving protocol namely, a periodic kick in the chemical potential to study the generation of Floquet MEMs in our setup. Our work paves a realistic way to engineer multiple MEMs in a driven system.
\end{abstract}

\maketitle

\section{Introduction}
Topological superconductors~(TSCs) hosting Majorana zero-modes~(MZMs) have been at the heart of modern condensed matter physics for the last two decades~\cite{Kitaev_2001,qi2011topological,Alicea_2012,Leijnse_2012,beenakker2013search,ramonaquado2017}. Majorana fermions are charge-neutral particles satisfying the Dirac equation, but unlike the Dirac fermions, they are self-antiparticles. However, in a condensed matter system, they do not constitute elementary particles rather they appear as emergent quasiparticles~\cite{Alicea_2012,beenakker2013search,Leijnse_2012,ramonaquado2017}. Majorana fermions obey non-Abelian statistics and are proposed to be the building block of fault-tolerant quantum 
computations~\cite{Ivanov2001,freedman2003topological,KITAEV20032,Stern2010,NayakRMP2008}. The first theoretical proposal for MZMs in a one-dimensional~(1D) topological system is put forward 
by Kitaev~\cite{Kitaev_2001}, based on a spinless $p$-wave superconducting chain. However, $p$-wave superconductors are not found naturally, and this restricts the experimental realization of the 
Kitaev's model. In this direction, Fu and Kane put forward the first experimentally feasible theoretical model proposal~\cite{FuPRL2008} based on topological insulators~(TIs). In their proposal, an $s$-wave superconductor is placed in closed proximity to the surface states of a three-dimensional~(3D) strong topological insulator and a magnetic insulator. An effective spiness $p_x+ip_y$ pairing is generated in the system due to the interplay between the $s$-wave superconductivity, the spin-orbit coupling~(SOC), and the exchange field. This system supports Majorana bound states at the vortex core. The generation of this spinless $p_x+ip_y$ superconductivity has also been described in other models~\cite{StanescuPRB2010,PotterPRB2011}. 

However, an alternative elegant proposal for realizing the MZMs is proposed in systems consisting of 1D semiconducting nanowirewire~(NW) (\eg InAs, InSb, etc.) with strong SOC and proximity induced $s$-wave superconductivity (\eg Nb) in it~\cite{SauPRL2010,LutchynPRL2010,Oreg2010,SauMZMPRB2010,TewariPRL2012,Leijnse_2012,Alicea_2012,ramonaquado2017,PhysRevB.87.024515,PhysRevB.86.180503,HaimPRL2015,BarmanPRB2021}. Such proposals, based on semiconductor-superconductor heterostructure, have attracted few recent experiments. The $2e^2/h$ quantized zero-bias peak, obtained via the tunneling spectroscopy measurements, has been reported as an indirect signature of the MZMs~\cite{das2012zero,Mourik2012Science,DengNano2012,Deng1557,NichelePRL2017,JunSciAdv2017,Zhang2017NatCommun,Gul2018,Grivnin2019,ChenPRL2019,PhysRevB.106.075306}. 
However, such quantization of the zero-bias peak may also arise due to the Andreev bound states induced by the quantum dots formed at the junction interface or the Kondo effect due to the magnetic impurities present in the system~\cite{KellsPRB2012,LiuPRL2012,LiuPRB2017,MoorePRB2018,VuikSciPost2019,LaiarXiv2021,doi:10.1126/science.abf1513}. 
Thus, the compelling and distinctive signature of the MZMs are yet to be found.

\begin{figure}[]
	\centering
	\subfigure{\includegraphics[width=0.4\textwidth]{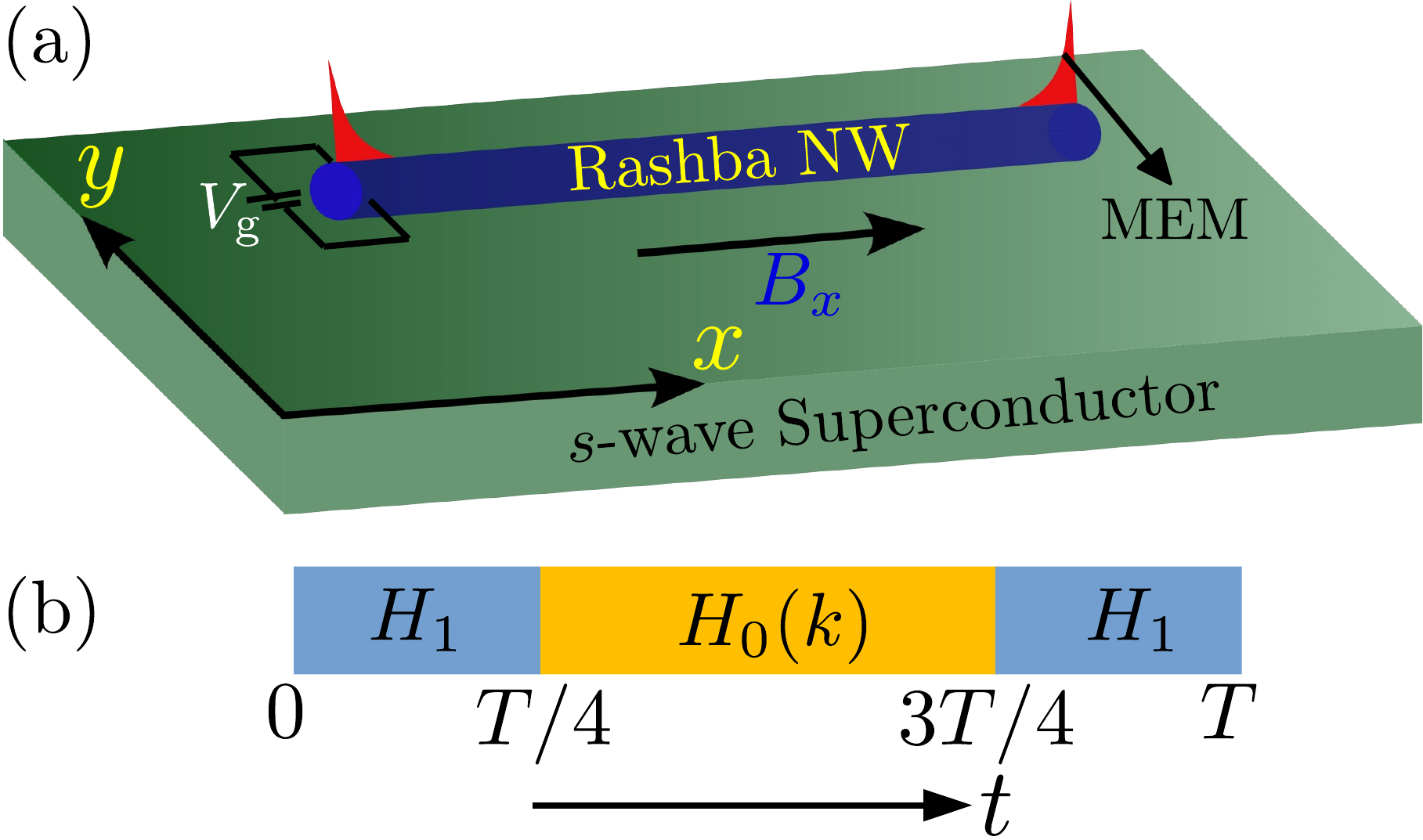}}
	\caption{(a) Schematic representation of our setup is shown here. One-dimensional nanowire~(NW) [blue] with strong spin-orbit coupling is placed on top a $s$-wave superconducting slab [green]. 
	A magnetic field $B_x$ is applied along the $x$-direction and a gate voltage $V_{\rm g}$ is applied across the cross-section of the NW to control the chemical potential. The Majorana end-modes~(MEMs) [red] appear at the two ends of the NW. (b) Schematic demonstration of our periodic three step-drive protocol~[see Eq.~(\ref{step_drive_protocol})] is presented.}
	\label{schematic}
\end{figure} 

On the other hand, the Floquet generation of topological systems has been capturing much attention due to its ability to engineer topological phenomena out of a non-topological system~\cite{oka09photovoltaic,kitagawa11transport,lindner11floquet,Rudner2013,Usaj2014,Piskunow2014,Eckardt2017,Yan2017,oka2019,NHLindner2020,nag2021anomalous,ThakurathiPRB2013,benito14,PotterPRX2016,
JiangPRL2011,ReynosoPRB2013,LiuPRL2013,ThakurathiPRB2017,MitraPRB2019,YangPRL2021,sacramento15,rxzhang21,PhysRevB.100.041103,PhysRevB.101.155417}. The non-trivial winding of time-dependent wave-function allows one to generate the anomalous topological boundary modes at finite energy, namely $\pi$-modes, which do not have any static counterpart. The Floquet generation of TSCs hosting Majorna end-modes~(MEMs), relying upon the spinless $p$-wave Kitaev chain, has been investigated~\cite{ThakurathiPRB2013,benito14,PotterPRX2016} along with their transport signature~\cite{KunduPRL2013}. To add more, 1D cold-atomic 
NW-$s$-wave superconductor heterostructure using the step-drive protocol~\cite{JiangPRL2011}, time-dependent ac electric field~\cite{ThakurathiPRB2017}, periodically driven chemical potential~\cite{LiuPRL2013,YangPRL2021,MitraPRB2019} are found to be instrumental in the context of Floquet TSCs~(FTSCs).  The  braiding of these Floquet modes further enrich the  field of quantum computations~\cite{Bomantara18,BomantaraPRB2018,BelaBauerPRB2019,MatthiesPRL2022}.
Importantly, following the realistic Rashba NW model, the generation of multiple MEMs similar to Ref.~\cite{ThakurathiPRB2013} has not been explored so far to the best of our knowledge. Also, these FTSCs hosting $0$- and $\pi$-MEMs have not been characterized using a proper dynamical topological invariant.
To this end, we ask the following questions- (a) Is it possible to generate the FTSC with multiple $0$- and $\pi$-MEMs stating from a static realistic model based on 1D Rashba NW? (b) How to characterize these dynamical modes using a proper dynamical invariant? (c) Are these Floquet MEMs stable against static random disorder? 
The interplay between Floquet physics and disorder might be useful in the context of topological characterization of the anomalous MEMs. We additionally note that the above questions are practically pertinent given 
the experimental developments of Floquet systems based on solid-state materials~\cite{WangScience2013,McIver2020}, acoustic setup~\cite{Peng2016,fleury2016floquet}, photonic platforms~\cite{RechtsmanExperiment2013,Maczewsky2017}, etc.

In this article, we first revisit the static model based on 1D Rashba NW in the presence of a magnetic field and proximity induced $s$-wave superconductivity (see Fig.~\ref{schematic})~\cite{SauPRL2010,LutchynPRL2010,Oreg2010,SauMZMPRB2010,TewariPRL2012,Leijnse_2012,Alicea_2012,HaimPRL2015,BarmanPRB2021} and study its topological phase boundaries (see Fig.~\ref{static}). We exploit the chiral symmetry to define the topological invariant for this system. Afterward, we tend towards the Floquet generation of TSC using a three-step drive protocol (see Fig.~\ref{stepeigen}). 
We employ the periodized evolution operator to topologically characterize the dynamical modes (see Fig.~\ref{stepPhase}). The robustness of the dynamical modes is also investigated against on-site random disorder (see Fig.~\ref{WindingDisorder}). In the presence of disorder, the translation symmetry no longer holds, hence we define a real-space topological invariant to characterize these dynamical modes (both $0$- and $\pi$-MEMs). We extend our analysis for the mass kick drive protocol to verify our proposals on Floquet Majorana modes (see Figs.~\ref{Kick}, and ~\ref{kickPhase}).   

The remainder of the article is organized as follows. We introduce the static Hamiltonian and its topological characterization in Sec.~\ref{Sec:II}. In Sec.~\ref{Sec:III}, we discuss the generation of Floquet MEMs and their topological characterization. In the presence of static random disorder, the stability of the dynamical MEMs via the appropriate topological invariant is studied in Sec.~\ref{Sec:IV}. 
We also introduce periodic kick protocol to generate the MEMs in Sec.~\ref{Sec:V}. We discuss the possible experimental connection and feasibility of our theoretical proposal in Sec.~\ref{Sec:experiment}.
We finally summarize and conclude our findings in Sec.~\ref{Sec:VI}.

\section{Model}\label{Sec:II}
\subsection{Model Hamiltonian}
We consider the Rashba NW model, placed on the top of a $s$-wave superconductor [see Fig.~\ref{schematic}~(a)], while a magnetic field is applied along the $x$-directon~\cite{Oreg2010}. The superconducting gap is induced in the NW via the proximity effect. We consider the following BdG basis: $\Psi_{k}=\{\psi_{k\ua},\psi_{k\da},\psi_{-k\da}^{\dagger},-\psi_{-k\ua}^{\dagger}\}^{\vect{t}}$; here, $\psi_{k \ua}~(\psi^\dagger_{k \ua})$ and $\psi_{k \da}~(\psi^\dagger_{k \da})$ represents electron annihilation~(creation) operator for spin-up and spin-down sector, respectively; and $\vect{t}$ stands for the transpose operation. The BdG Hamiltonian for the NW is given  as~\cite{LutchynPRL2010,Oreg2010,TewariPRL2012,Leijnse_2012,Alicea_2012,HaimPRL2015}
\begin{equation}
H_{0}(k) = [(2t-c_0)-2t \cos k] \Gamma_1 + 2u \sin k \Gamma_2 + B_x \Gamma_3 + \Delta \Gamma_4 \
\label{eq:static_Hamiltonian},
\end{equation}
where, the $4\times 4$ $\vect{\Gamma}$ matrices are given as $\Gamma_{1}=\tau_{z} \sigma_{0}$, $\Gamma_{2}=\tau_{z} \sigma_{z}$, $\Gamma_{3}=\tau_{0} \sigma_{x}$, and $\Gamma_{4}=\tau_{x} \sigma_{0}$; while the Pauli matrices $\vect{\tau}$ and $\vect{\sigma}$ act on particle-hole and spin ($\uparrow$, $\downarrow$) subspace, respectively. Here, $c_{0}$, $t$, $u$, $B_x$, and $\Delta$ represent the chemical potential, nearest-neighbor hopping amplitude, Rashba SOC strength, strength of the magnetic field along $x$-direction, and induced $s$-wave superconducting gap, respectively. The Hamiltonian [Eq.~(\ref{eq:static_Hamiltonian})] respects the chiral symmetry $S=\tau_{y}\sigma_{z}$: $S^{-1} H_{0}(k) S = - H_{0}(k)$ and the particle-hole symmetry~(PHS) $\mathcal{C}=\tau_y \sigma_y \mathcal{K}$: $\mathcal{C}^{-1} H_{0}(k) \mathcal{C} = - H_{0}(-k)$. However, it breaks the time-reversal symmetry~(TRS) $\mathcal{T}=i \tau_0 \sigma_y \mathcal{K}$: $\mathcal{T}^{-1} H_{0}(k) \mathcal{T} \neq H_{0}(-k)$; where, $\mathcal{K}$ represents the complex-conjugation operator.

\begin{figure}[]
	\centering
	\subfigure{\includegraphics[width=0.48\textwidth]{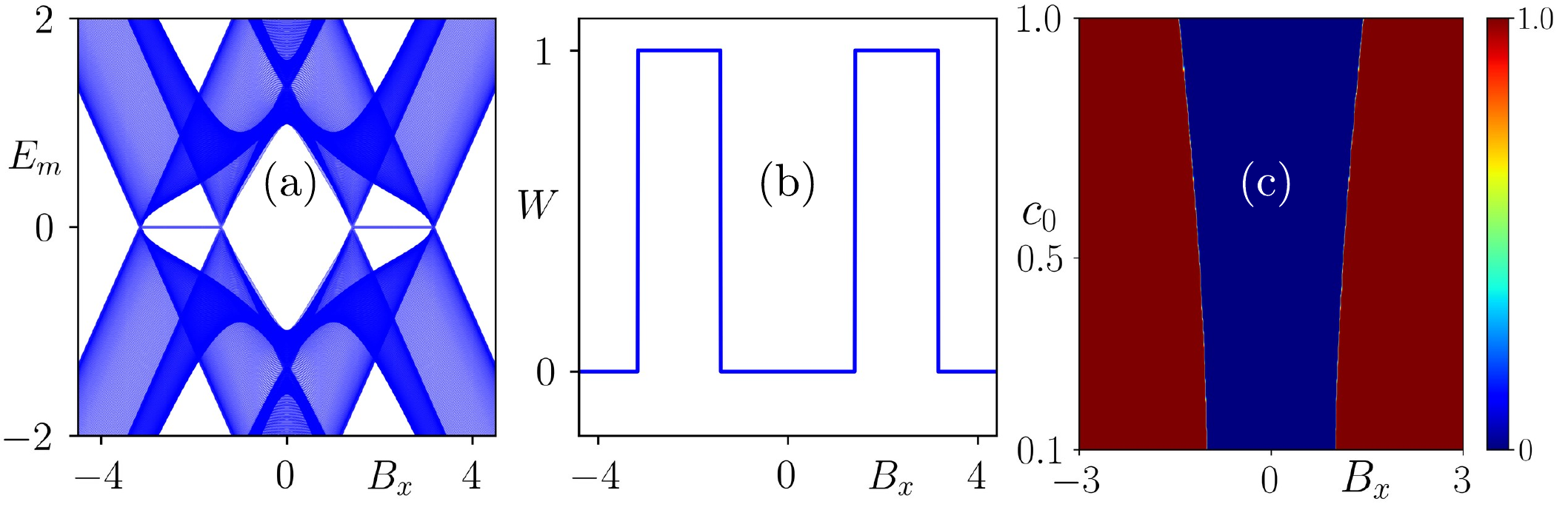}}
	\caption{(a) Energy eigenvalue spectra for the static Hamiltonian [Eq.~(\ref{eq:static_Hamiltonian})], employing open boundary condition, is shown as a function of the magnetic field $B_{x}$. The corresponing winding number $W$ is depicted as a function of $B_{x}$ in panel (b) for a fixed value of $c_0=1.0$. (c) We demonstrate $W$ in the $c_{0} \mhyphen B_{x}$ plane to illustrate the topological phase diagram. Other model parameters are chosen as $(t, u,\Delta)=(1.0, 0.5, 1.0)$. MZMs are obtained for $\lvert B_x^{1} \rvert \le B_x \le \lvert B_x^{2} \rvert$ (see text for details).
	}
	\label{static}
\end{figure} 

At $k=0$, the Hamiltonian~[Eq.~(\ref{eq:static_Hamiltonian})] undergoes a gap-closing transition, when $ B_{x}=\lvert B_x^{1} \rvert= \sqrt{c_{0}^2 + \Delta^2}$; while another gap-closing transition occurs at $k=\pi$, when $B_{x}=\lvert B_x^{2} \rvert= \sqrt{(4t-c_0)^2+\Delta^2}$. These bulk gap closing conditions can identify the topological phase boundaries. Having emphasized the problem analytically, we tie up with numerical results to corroborate our findings. We employ open boundary condition~(OBC) and depict the eigenvalue spectra of the Hamiltonian [Eq.~(\ref{eq:static_Hamiltonian})]
in Fig.~\ref{static}~(a) as a function of the magnetic field $B_x$. From Fig.~\ref{static}~(a), one can identify that the system is trivially gapped when $B_x < \lvert B_x^{1} \rvert$ and $B_x > \lvert B_x^{2} \rvert$ with no MZMs present. However, a pair of MZMs appear at the two ends of NW~(one MZM per end), when $\lvert B_x^{1} \rvert \le B_x \le \lvert B_x^{2} \rvert$ identifying the TSC phase in the 
static model. 

\subsection{Topological characterization}
The 1D system under consideration preserves chiral symmetry $S$. Thus, it can be topologically characterized by a $\mathbb{Z}$ winding number, classifying Hamiltonian's first homotopy class within the first homotopy group $\Pi_1[U(N)]$ and providing us with the number of zero-energy modes~\cite{RyuNJP2010,ChiuRMP2016}. In the canonical basis representation (chiral-basis), where $S$ is diagonal, the Hamiltonian [Eq.~(\ref{eq:static_Hamiltonian})] takes an anti-diagonal form. This reads as
\begin{equation}
\tilde{H}_{0}(k)=U_S^{\dagger}H_{0}(k)U_S=
\begin{pmatrix}
0&H_{0}^{+}(k)\\
H_{0}^{-}(k)&0
\end{pmatrix}
\label{eq:anti_static_Hamiltonian},
\end{equation}
where, the unitary matrix $U_S$ is constructed using the chiral-basis, given as
\begin{equation}
U_S=\frac{1}{\sqrt{2}}
\begin{pmatrix*}[r]
0&i&0&-i\\
-i&0&i&0\\
0&1&0&1\\
1&0&1&0
\end{pmatrix*} \
\label{chiral_basis} ,
\end{equation}
and $H_{0}^{\pm}(k)$ are $2\times2$ square matrices, defined on the $\pm$ chiral block, respectively. Using ${\cal H}_{0}^{\pm}(k)$, we can define the winding number $W$~$(\in \mathbb{Z})$ as~\cite{RyuNJP2010,ChiuRMP2016} 
\begin{equation}
W=\Bigg| \pm \frac{i}{2 \pi} \int_{- \pi}^{\pi} dk \hspace{0.1 cm} {\rm Tr} \left[\left\{{ H}_{0}^{\pm}(k)\right\}^{-1} \partial_{k} { H}_{0}^{\pm}(k)\right] \Bigg|
\label{eq:static_wind}.
\end{equation}
We illustrate $W$ as a function of $B_x$ for a fixed value of $c_0$ in Fig.~\ref{static}~(b). The winding number $W$ correctly identifies the phase boundary between the trivial $W=0$ and the 
topological $W=1$ phases. 
In order to obtain the topological phase diagram, we further depict the winding number in the $c_0 \mhyphen B_x$ plane in Fig.~\ref{static}~(c). The system exhibits TSC phase with $W=1$ for 
$B_x>| B_x^1 |$. Our findings are consistent with the previously reported results~\cite{Oreg2010,Leijnse_2012,Alicea_2012} and constitute the background for the investigation of the driven system 
as discussed in the next section.

\section{
Floquet Majorana modes and their Topological Characterization}\label{Sec:III}

\subsection{Driving protocol and Emergence of Floquet Majorana modes}

\begin{figure}[]
	\centering
	\subfigure{\includegraphics[width=0.48\textwidth]{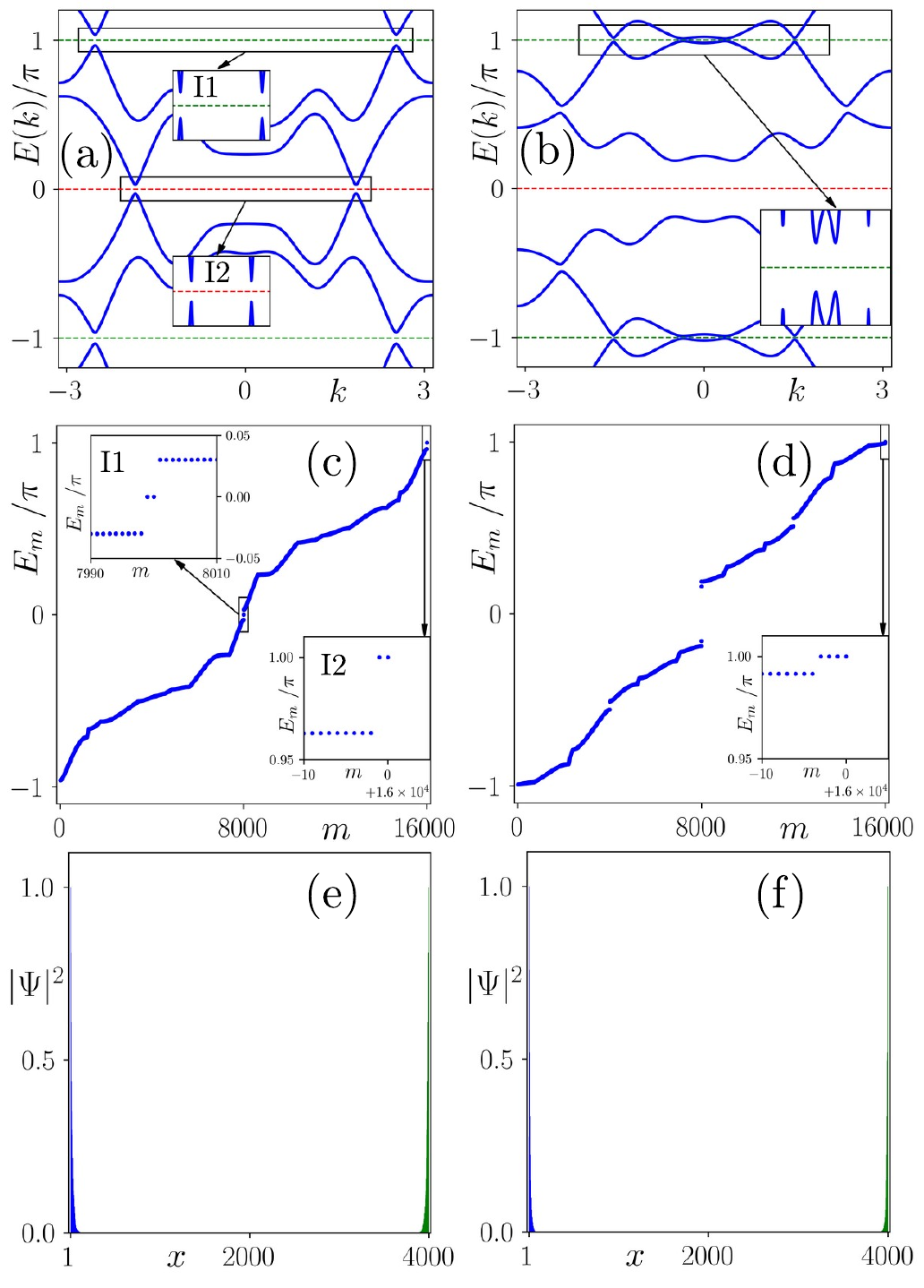}}
	\caption{Bulk quasienergy spectra for the three-step drive protocol [Eq.~({\ref{step_drive_protocol}})] are shown for $\Omega = 1.5$ and $\Omega = 2.5$ in panels (a) and (b), respectively. In the insets I1 and I2 of panel (a)~[inset of panel (b)], we depict the bulk-gap around quasi-energy $E(k)=\pi$ and $E(k)=0$, respectively~[quasi-energy $E(k)=\pi$]. In panels (c) and (d), we illustrate the quasi-energy spectra $E_m$ as a function of the state index $m$ for our system obeying OBC, corresponding to (a) and (b), respectively; while in the insets, we portray the zoomed-in quasi-energies for better clarity. The appearance of $0$- and $\pi$-Majorana end modes~(MEMs) are evident from these figures. We show the normalized site-resolved probability~($\lvert \Psi \rvert^2$) corresponding to the $0$- and $\pi$-MEMs in panels (e) and (f), respectively. Here, we consider $N=4000$ lattice sites to obtain sharp Majorana localization at individual ends of the NW. All the other model parameters are chosen as $(c_{0},c_{1},\Delta,B_{x}, t, u)= (1.0, 0.2, 1.0, 1.0, 1.0, 0.5)$.
	}
	\label{stepeigen}
\end{figure}

We choose a non-topological parameter space for the Hamiltonian $H_0(k)$ [Eq.~(\ref{eq:static_Hamiltonian})] to start with and consider a three-step periodic driving scheme to generate 
the FTSC phase as
\begin{eqnarray}
{H}_{\rm step}(k,t) && = H_{1} \hspace*{2.5 cm} t \in \left[0,T/4\right) \ , \nonumber \\
&& = { H}_{0}(k) \hspace*{2.0 cm} t \in \left[T/4,3T/4\right) \ , \nonumber \\
&& = H_{1} \hspace*{2.5 cm} t \in \left[3T/4,T\right] \
\label{step_drive_protocol},
\end{eqnarray}
where, $H_{1}=-c_{1} \Gamma_{1}$ represents an on-site term, modulating the chemical potential. Here, $T$~($\Omega$) stands for the time-period~(frequency) of the drive. We depict the driving protocol schematically in Fig.~\ref{schematic}(b). The Floquet operator can be constructed as follows 
\begin{eqnarray}
U_{\rm step}(k,T) &=& {\rm TO} \exp \left[-i \int_{0}^{T} dt \  H(k,t)  \right] \non \\
&=&e^{-i H_{1}T/4 } e^{-i H_{0}(k)T/2} e^{-i H_{1}T/4 } \
\label{step_evolution_operator}. 
\end{eqnarray}
We can diagonalize the Floquet operator $U_{\rm step}(k,T)$ to obtain the quasi-energy spectrum as $U_{\rm step}(k,T) \ket{\Psi_n(k)}=e^{-i E'_n(k) T} \ket{\Psi_n(k)}$. Here, $n$ represents the band index in the quasi-energy spectrum. We define $E_n(k)=E_n'(k)T$ such that the quasi-energy always lies within the range \ie $E_n(k) \in \left[-\pi,\pi\right]$.

We depict the bulk quasi-energy bands $E(k)$ as a function of momenta $k$ for $\Omega=1.5$ and $2.5$ in Fig.~\ref{stepeigen}~(a) and (b), respectively. Although, the  bulk quasi-energy bands do not exhibit any direct signature of the end-modes due to the periodic boundary condition. However, by analyzing the bulk bands, one can obtain a qualitative idea about the number of end-modes present 
within a certain quasi-energy gap. From the insets (I1 and I2) of Fig.~\ref{stepeigen}~(a), we can observe two openings of the bulk gap corresponding to two MEMs at these gaps. In these insets, $0$- and 
$\pi$-lines are indicated by the red and green dotted lines, respectively. Similarly, from the inset of Fig.~\ref{stepeigen}~(b), one can identify four gap openings near $E(k)=\pi$ (denoted by the green dotted line), corresponding to four end-modes at that gap. Nonetheless, we also verify our claims from the finite geometry calculations employing OBC. 

Having studied the bulk quasi-energy bands, we impose OBC and diagonalize the Floquet operator [Eq.~(\ref{step_evolution_operator})]. We demonstrate the quasi-energy eigenvalue spectra for 
$\Omega=1.5$ and $\Omega=2.5$ in Fig.~\ref{stepeigen}~(c) and (d), respectively. We can identify two $0$-MEMs (one $0$-Majorana mode per end) and two $\pi$-MEMs (one $\pi$-Majorana per end) from the insets I1 and I2 of Fig.~\ref{stepeigen}~(c), respectively. From Fig.~\ref{stepeigen}~(d), however, we can observe only four $\pi$-MEMs (two $\pi$-Majoranas per end), while $0$-MEMs are absent in this case. This is also emphasized in the inset of Fig.~\ref{stepeigen}~(d). Thus, the finite geometry calculations verify the findings as anticipated from the bulk quasi-energy bands. The information about the nature of the bulk gap whether it is topological or trivial is acquired by investigating the existence of MEMs within the bulk gap under OBC. The winding number captures the topological property of the bulk gap that we demonstrate below.   

To study the localization properties of the in-gap states, we illustrate the normalized site-resolved probability~($\lvert \Psi \rvert^2$) in Fig.~\ref{stepeigen}~(e) and (f) corresponding to the $0$- and $\pi$-MEMs as shown in Fig.~\ref{stepeigen}~(c), respectively. One can observe that both the $0$- and $\pi$-MEMs are sharply localized near the two ends of the NW. To corroborate that the end-modes represent a Majorana, we consider the wavefunction of the $0$- and $\pi$-modes, corresponding to Fig.~\ref{stepeigen}~(e) and (f), respectively. For the $i$$^{\rm th}$ lattice site, the wavefunction takes the form $\Psi_{i}=\{\psi_{i\ua},\psi_{i\da},\psi_{i\da}^{\dagger},-\psi_{i\ua}^{\dagger}\}^{\vect{t}}$. We further verify that $|\psi_{i\ua}|=|\psi_{i\ua}^{\dagger}|$ and $|\psi_{i\da}|=|\psi_{i\da}^{\dagger}|$ for both the $0$- and $\pi$-modes. Hence, both the $0$- and $\pi$-modes represent Majorana modes being their own anti-particle. Note that, the amplitude of the electron and hole part is found to be identical  for a given species of spin. This refers to the emergent spinless $p$-wave nature of the supercoducting gap within which MEMs exist. 

\subsection{Topological characterization}
We exploit the chiral symmetry to characterize the topological properties of dynamical MEMs. The $0$- and $\pi$-MEMs cannot be characterized distinctly using the Floquet operator only. To resolve this, we introduce the notion of gap $\epsilon$ via the periodized evolution operator as~\cite{Rudner2013,GhoshPRB2022}
\begin{equation}
U_{\epsilon}(k,t)= U_{\rm step}(k,t) \left[U_{\rm step}(k,T)\right]_{\epsilon}^{-t/T} \ ,
\label{periodise_evolution_operator}
\end{equation}
where, $U_{\rm step}(k,t)$ represents the time-evolution operator and $(-t/T)$$^{\rm th}$ power of the Floquet operator for the $\epsilon$-gap $\left[U_{\rm step}(k,T)\right]_{\epsilon}^{-t/T}$ is given as
\begin{eqnarray}
\left[U_{\rm step}(k,T)\right]_{\epsilon}^{-t/T}&=&  \sum_{n=1}^{N/2} e^{-i(2\epsilon-|E_{n}(k)|)t/T} \ket{\psi_n(k)} \bra{\psi_n(k)} \nonumber \\
 &&\hspace{-0.8 cm}+ \sum_{n=N/2 +1}^{N} e^{-i \left|E_{n}(k) \right|t/T} \ket{\psi_n(k)} \bra{\psi_n(k)}
 \label{uzero}, \quad 
\end{eqnarray}
where, $n=1 \cdots N/2$ ($N/2+1 \cdots N$) represent vanlence~(conduction) bands. Note that, $U_{\epsilon}(k,0)=U_{\epsilon}(k,T)=\mathbb{I}$ and $U_{\epsilon}(k,t)$ is periodic in time \ie $U_{\epsilon}(k,t)=U_{\epsilon}(k,t+T)$. Since, $H_{\rm step}(k,t)$ respects chiral symmetry, one can show that the chiral symmetry $S$ imposes the following constraints on the periodized evolution operator~\cite{Michel2016} as
\begin{equation}\label{constraintsPEO}
	S U_{\epsilon}(k,t) S^{-1}= U_{-\epsilon}(k,-t) e^{2 \pi i t /T} \ . 
\end{equation} 
Using the periodicity condition of $U_{\epsilon}(k,t)$, we obtain at the half period: $S U_{\epsilon}\left(k,\frac{T}{2}\right) S^{-1}=-U_{-\epsilon}\left(k,\frac{T}{2}\right)$. At the half-period $t=\frac{T}{2}$, we obtain the constraints on the $U_{\epsilon}\left(k,\frac{T}{2}\right)$ for $0$- and $\pi$-gap as
\begin{eqnarray}
	S U_{0}\left(k,\frac{T}{2}\right) S^{-1}&=&-U_{0}\left(k,\frac{T}{2}\right) \non \ , \\
	S U_{\pi}\left(k,\frac{T}{2}\right) S^{-1}&=&U_{\pi}\left(k,\frac{T}{2}\right) \ ,
\end{eqnarray} 
where, we have used the relation $U_{-\pi}\left(k,\frac{T}{2}\right)=-U_{\pi}\left(k,\frac{T}{2}\right)$. Thus, in the chiral basis, $U_{0}\left(k,\frac{T}{2}\right)$~$\left[U_{\pi}\left(k,\frac{T}{2}\right)\right]$ takes block anti-diagonal [diagonal] form. This reads as
\begin{eqnarray}
\tilde{U}_{0}\left(k,\frac{T}{2}\right)&=&U_S^{\dagger} U_{0}\left(k,\frac{T}{2}\right) U_S = 
\begin{pmatrix}
0& U_{0}^{+}(k)\\
U_{0}^{-}(k)&0
\end{pmatrix} , \  \ \quad
\label{step_antidiag_U_zero} \\
\tilde{U}_{\pi}\left(k,\frac{T}{2}\right)&=&U_S^{\dagger} U_{\pi}\left(k,\frac{T}{2}\right) U_S = 
\begin{pmatrix}
	U_{\pi}^{+}(k)&0\\
	0&U_{\pi}^{-}(k)
\end{pmatrix}
\label{step_diag_U_pi}.
\end{eqnarray}
Using $U_{\epsilon}^{\pm}(k)$, the dynamical winding number $W_{\epsilon}$ for $\epsilon(=0,\pi)$-gap can be defined as~\cite{Michel2016}
\begin{equation}
W_{\epsilon}= \left| \pm \frac{i}{2 \pi} \int_{-\pi}^{\pi} dk \ {\rm Tr} \left[\left\{U_{\epsilon}^{\pm}(k)\right\}^{-1} \partial_{k}  U_{\epsilon}^{\pm}(k)\right] \right|
\label{dynamical_wind_step}.
\end{equation}
Here, $W_{\epsilon}$ counts the number of modes present per end of the NW at the $\epsilon$-gap. Thus, $W_{\epsilon}$ serves as the dynamical analog of Eq.~(\ref{eq:static_wind}).

\begin{figure}[]
	\centering
	\subfigure{\includegraphics[width=0.48\textwidth]{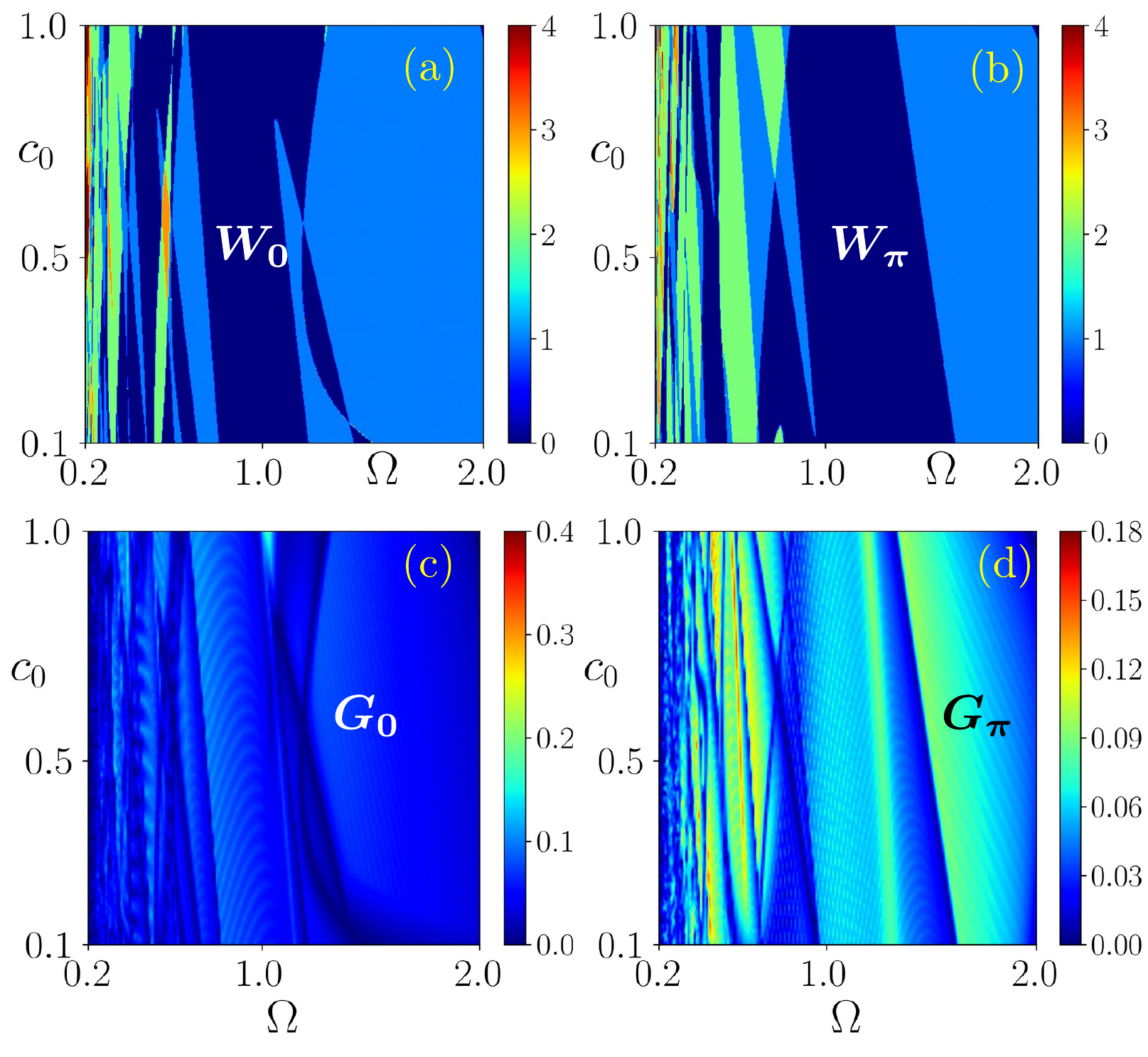}}
	\caption{We demonstrate the dynamical winding number $W_0$ and $W_\pi$ (quasi-energy gap corresponding to $0$-energy, $G_{0}$ and $\pi$-energy, $G_{\pi}$) in $c_0 \mhyphen \Omega$ plane in panels (a) and (b) ((c) and (d)), respectively. Here the color bar represents the winding number (energy gap) for (a), (b) ((c), (d)). All the other parameters are chosen to be the same as mentioned in Fig.~\ref{stepeigen}.
	}
	\label{stepPhase}
\end{figure}
We demonstrate the topological phase diagram in terms of $W_{0}$ and $W_{\pi}$ in the $c_{0}$-$\Omega$ plane in Fig.~\ref{stepPhase} (a) and (b), respectively. The color bar represents the number 
of MEMs presents in the system. One can observe that the numbers of both $0$- and $\pi$-MEMs are found  to be more than one in certain parameter regime at low frequency. We find that, in the intermediate range of frequency $\Omega \in[0.35, 1.0]$ (while hopping parameter is set to unity), one can observe various topological phases with multiple Majorana $0$- and $\pi$-modes. Therefore, from the experimental point of view such an low-intermediate range of frequency might be useful for obtaining multiple $0$- and $\pi$-MEMs.  In the case of lower frequency $\Omega <0.35$, the topological phases become short lived in the parameter space resulting in a problem for choosing a suitable experimentally feasible parameter range. On the other hand, intermediate to moderately high frequency range ($\Omega\sim 1.0 - 2.0$) can also be suitable in terms of engineering single Majorana $0$- and $\pi$-modes. 

In Fig.~\ref{stepPhase}(c) and (d), we show the quasienergy gap around the quasienergy $0$ and $\pi$, namely $G_0$ and $G_{\pi}$, respectively. The different phases, characterized by a well-defined dynamical winding number, are gapped out, while on the phase-boundaries, associated with vanishing gaps $G_{0,\pi}=0$, dynamical winding numbers are ill-defined. Therefore, there exists a one-to-one correspondence between the gap and the invariant. Note that the system is in the trivial phase when the winding number becomes zero. One can analytically understand the $0$-energy gap in the high frequency $T\to 0$ limit from an effective Hamiltonian $H_{\rm eff}=H_0/2 + H_1/2$ that can be obtained from Brillouin-Wigner approximation~\cite{Mikami16}. As a result, the gap is modified by a parameter $c_1$ in addition to the parameters present in $H_0$.

The modulation in the MEM number is possible only in a driven system via the procreation of longer-range hoppings that is more probable at low frequency regimes. By contrast, high frequency drive 
is not able to generate higher-order hoppings resulting in one MEM similar to the static Hamiltonian. In a driven system, one can engineer more gap-closing transitions at quasi-energy $E(k)=0/\pi$~[see Fig.~\ref{stepeigen}~(b)], which in turn can give rise to the emergence of more Floquet MEMs at $E(k)=0/\pi$. The number of MEMs, residing at $0$ gap, can be different from that of at $\pi$ gap 
causing a rich Floquet topological phase diagram unlike the static case. The multiple Floquet MEMs are protected by the chiral symmetry. Generation of multiple $0$- and $\pi$-MEMs and their 
topological characterization employing appropriate dynamical invariant serves as the prime result of the current manuscript.

\section{Stability of Floquet MEMs in presence of disorder and their topological charecterization}\label{Sec:IV}
After investigating the generation and characterization of the Floquet $0$- and $\pi$-MEMs, we further explore the stability of these modes in the presence of random disorder. We consider the following on-site time-independent disorder in $H_{\rm step}(k,t)$ [Eq.~(\ref{step_drive_protocol})] as
\begin{equation}
	V_{\rm dis} = V(r) \Gamma_1 \ ,
\end{equation}
where, $V(r)$ is distributed randomly in the range $\left[-\frac{w}{2},\frac{w}{2}\right]$ and $w$ accounts for the disorder strength. Note that the system continues to preserve the chiral symmetry in the presence of the above disorder \ie $S^{-1} H(r) S= -H(r)$ where $H(r)$ denotes the real space Hamiltonian as $k$ is no longer a good quantum number. The  quasi-energy spectra remain qualitatively similar to Fig.~\ref{stepeigen}~(c)-(d) and continue to manifest both $0$- and $\pi$-modes as long as the disorder scale does not close these corresponding gaps (see text for further discussion).

\begin{figure}[]
	\centering
	\subfigure{\includegraphics[width=0.48\textwidth]{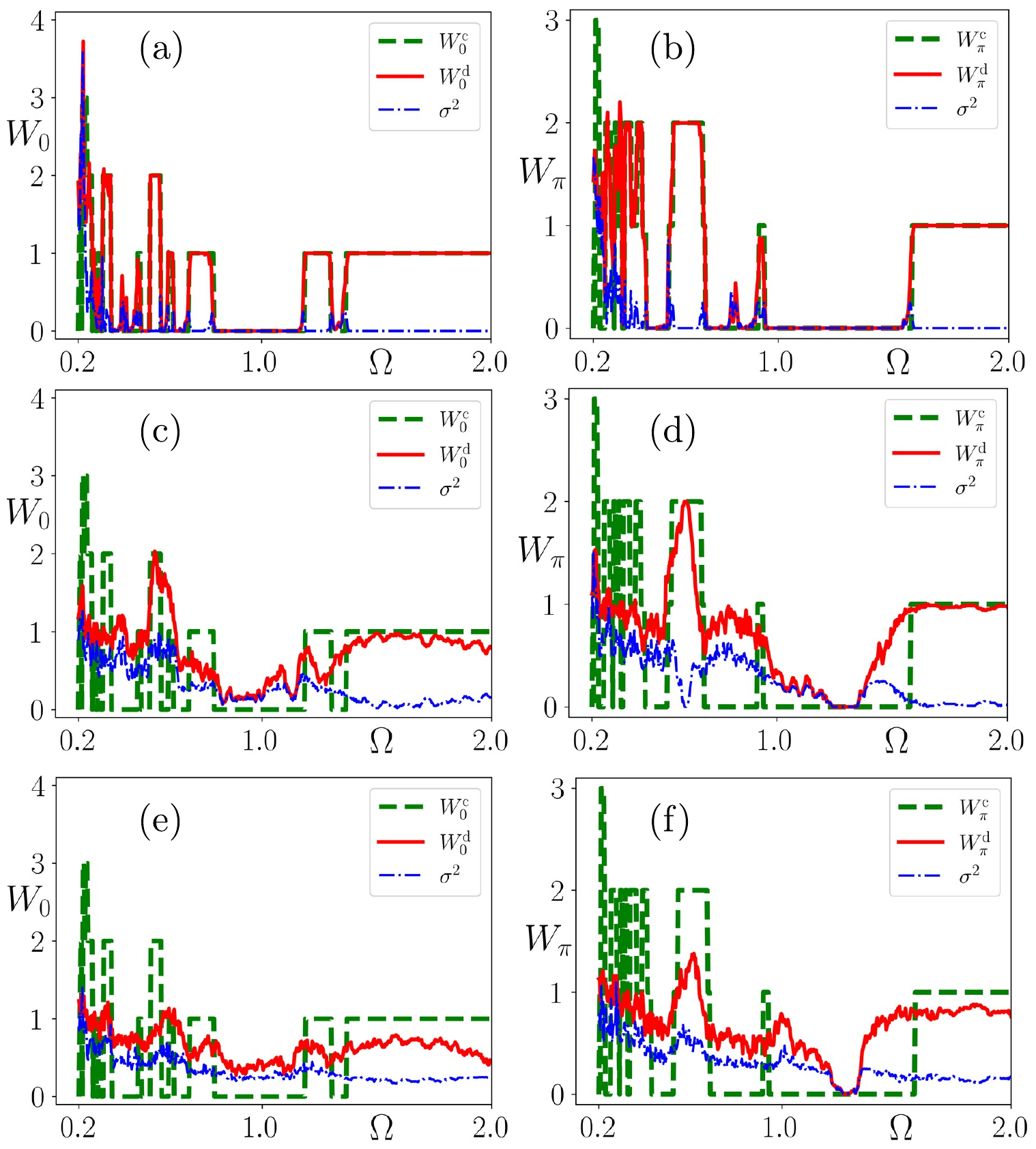}}
	\caption{In panel (a)-(b), we depict the dynamical winding number $W_0$ and $W_\pi$ respectively in presence of small $(w=0.1)$ onsite-disorder as a function of the driving frequency $\Omega$. We repeat (a)-(b) in panel (c)-(d) and (e)-(f) for moderate ($w=0.5$) and strong ($w=0.8$) disorder strength, respectively. Here, green (dashed), red (solid), and blue (dotted-dashed) lines represent the winding number ($W^{\rm c}_\epsilon$) for the corresponding clean system, disorder averaged winding number ($W^{\rm d}_\epsilon$) for the disorderd system, and the variance~($\sigma^2$) of winding number of the disordered system from the mean value $W^{\rm d}_\epsilon$, respectively for quasi-energy $\epsilon$. We choose $c_0=0.2$, while all the other model parameter values remain the same 
as mentioned in Fig.~\ref{stepeigen}.
	}
	\label{WindingDisorder}
\end{figure}

Due to the absence of quasi-momenta $k$, we cannot compute the dynamical winding number $W_\epsilon$ using Eq.~(\ref{dynamical_wind_step}) for the disordered system. However, we can impose twisted boundary condition~(TBC) to define the topological invariant~\cite{Demler2010PRB,Titum2016PRX,Sreejith2016} for this case. In the TBC, the two ends of the system are glued together to form a ring, and a periodic flux $\theta$ is threaded through the ring such that whenever a particle hops from one lattice site to another, it acquires a phase~\cite{Niu1984,NiuPRB1985}. This accounts for the following transformation: $\psi_{j,  \ua (\da)} \rightarrow e^{i \theta j }\psi_{j,\theta \ua (\da)}$ and $\psi_{j,  \ua (\da)}^\dagger \rightarrow e^{-i \theta j }\psi_{j,\theta \ua (\da)}^\dagger$ such that 
$L \theta=2 \pi$, where $L$ being the number of lattice sites. The lattice Hamiltonian in the presence of on-site disorder and the periodic flux $\theta$ can be written in the form
\begin{widetext}
\begin{eqnarray}\label{DrivenHamReal}
	H_{\rm step}&=& \sum_{j,\theta} \Psi_{j,\theta}^\dagger \Big[-c_1+V(r)\Big] \Gamma_1 ~ \Psi_{j,\theta}, \hspace{9.2cm} t \in \left[0,T/4\right)\non \\
	&=& \sum_{j,\theta} \Psi_{j,\theta}^\dagger \big[\{(2t-c_0)+V(r)\}\Gamma_1 +B_x \Gamma_3 + \Delta \Gamma_4 \big] \Psi_{j,\theta}+\Psi_{j,\theta}^\dagger (-t \Gamma_1 -i u \Gamma_2 ) e^{i \theta} \Psi_{j+1,\theta}  +\  {\rm h.c.} , ~~~ t \in \left[T/4,3T/4\right) \non \\
	&=&\sum_{j,\theta} \Psi_{j,\theta}^\dagger \Big[-c_1+V(r)\Big] \Gamma_1 ~ \Psi_{j,\theta}, \hspace{9.2cm} t \in \left[3T/4,T\right].\qquad
\end{eqnarray}	
\vskip -0.5cm
\end{widetext}
here, $\Psi_{j,\theta}=\{\psi_{j,\theta \ua},\psi_{j,\theta \da},\psi_{j,\theta \da}^{\dagger},-\psi_{j,\theta \ua}^{\dagger}\}^{\vect{t}}$ and $\theta$ serves the purpose of the momenta, whereas the unit cell in this superlattice is composed of $4L$ disordered-lattice.  
Note that, our analysis is independent of the choice of the gauge to represent the twist variable $\theta$. The chiral symmetry operator in the real space can be written as $S^L=S \otimes \mathbb{I}_L$. 
The $4L \times 4L$ chiral basis can be obtained by diagonalizing $S^L$ and using the chiral basis along with 
the relations Eqs.~(\ref{periodise_evolution_operator})-(\ref{step_diag_U_pi}), we can evaluate $U_{\epsilon}^{\pm}(\theta)$. However, $U_{\epsilon}^{\pm}(\theta)$ are now $2L \times 2L$ matrices. Incorporating these, we can define the dynamical winding number $W^{\rm d}_\epsilon$ for the disordered system as~\cite{Demler2010PRB,Titum2016PRX,Sreejith2016}
\begin{equation}
	W_{\epsilon}^{\rm d}= \left| \pm \frac{i}{2 \pi} \int_{-\pi}^{\pi} d\theta \ {\rm Tr} \left[\left\{U_{\epsilon}^{\pm}(\theta)\right\}^{-1} \partial_{\theta}  U_{\epsilon}^{\pm}(\theta)\right] \right|
	\label{dynamical_wind_step_disorder}.
\end{equation}
We exhibit disorder-averaged winding number $W_\epsilon^{\rm d}$, its variance $\sigma^2$ from the mean value, and winding number of the clean system $W_\epsilon^{\rm c}$ (for comparision) as a function of the driving frequency $\Omega$ in Fig.~\ref{WindingDisorder}. We have considered a system consisting of $L=30$ sites and performed an average over $100$ disorder configurations. 
We consider three sets of the disorder strenght- small ($w=0.1$), moderate ($w=0.5$), and strong ($w=0.8$) and discuss their outcome below: 

Case I - Small disorder ($w=0.1$): We show $W_{0}$ and $W_{\pi}$ in Fig.~\ref{WindingDisorder}~(a) and (b), respectively. The disorder-averaged winding number $W^{\rm d}_\epsilon$ matches well 
with that of the clean case $W^{\rm c}_\epsilon$ except near smaller $\Omega$ values. The variance $\sigma^2$ remains almost near zero, and also manifests the stability of $W^{\rm d}_\epsilon$. 
However, the finite variance indicates the fluctuations in the quantization in low frequency regime. This discrepancy can be attributed to the smaller bulk gaps near $0$ and $\pi$ quasi-energy.

Case II - Moderate disorder ($w=0.5$): We show $W_{0}$ and $W_{\pi}$ in Fig.~\ref{WindingDisorder}~(c) and (d), respectively. However, $W^{\rm d}_\epsilon$ matches with that of the clean case $W^{\rm c}_\epsilon$ only near higher values of $\Omega$ (high-frequency limit). The variance $\sigma^2$ also exibits large fluctuations near smaller frequencies and tends to zero near higher frequencies. Thus MEMs survive for relatively higher frequency regime even in presence of moderate disorder strength. More interestingly, the Majorana $\pi$-modes  appear to be more robust against disorder (near $\Omega\sim 0.5$) as compared to the Majorana $0$-modes as the former deviates less from the quantized value (the variance $\sigma^2\sim 0$) of the corresponding winding number 
(see Figs. 5 (c) and (d)). This can be attributed to the fact that disorder usually affects the Floquet Majorana modes when its energy scale is grater or equivallent to the topological gap. 
For moderate disorder, the corresponding scale may not be comparable to the $\pi$-gap while it can be commensurate to close the $0$-gap in the observed frequency limit. Hence, $\pi$-modes appear 
to be more robust against disorder as compared to the Majorana $0$-modes.

Case III - Strong disorder ($w=0.8$): We show $W_{0}$ and $W_{\pi}$ in Fig.~\ref{WindingDisorder}~(e) and (f), respectively. There is almost no matching between $W^{\rm d}_\epsilon$  and $W^{\rm c}_\epsilon$. The variance $\sigma^2$ is non-zero throughout the frequency range. Both the $0$- and $\pi$-MEMs disappear for strong disorder strength.

\section{Generation of Floquet MEMs using Periodic Kick drive protocol
}\label{Sec:V}

\begin{figure}[]
	\centering
	\subfigure{\includegraphics[width=0.48\textwidth]{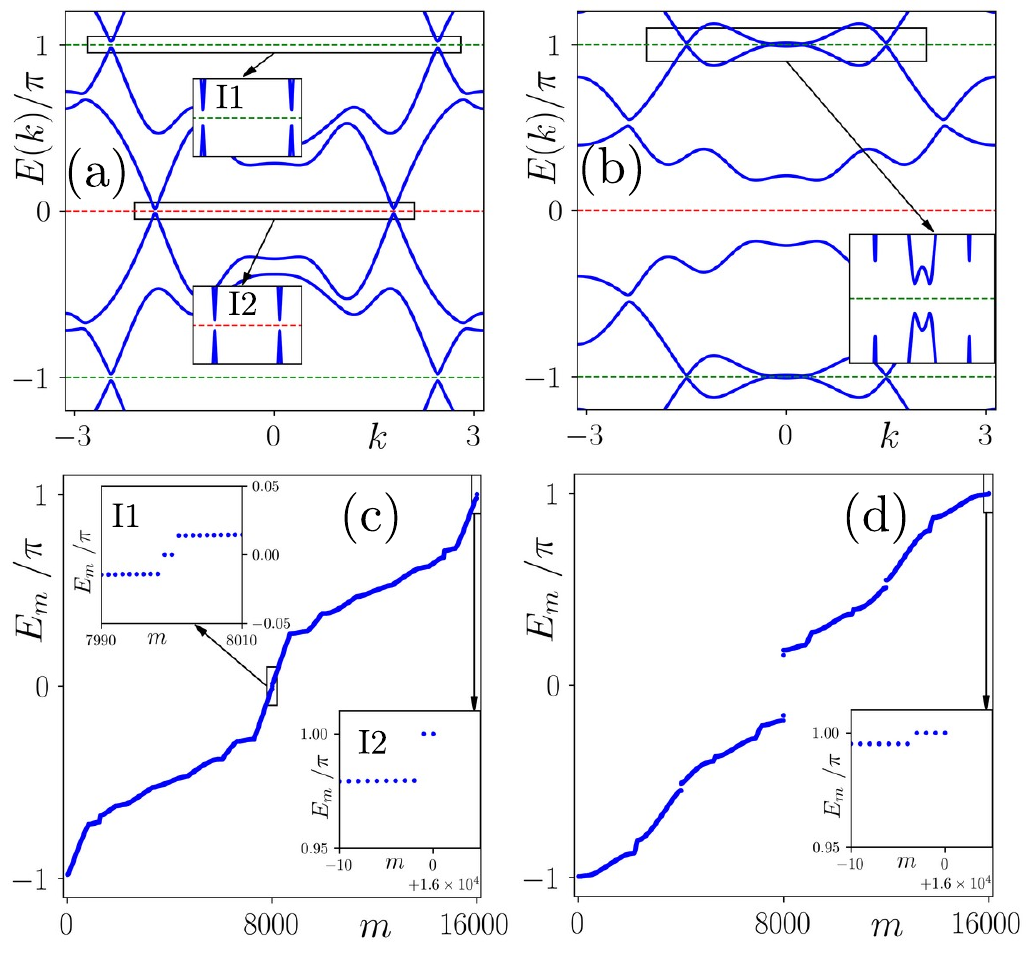}}
	\caption{(a) and (b) panels represent the bulk quasi-energy spectra for the periodic-kick drive protocol [Eq.~(\ref{kick_drive_protocol})] for $\Omega=3.0$ and $\Omega=5.0$, respectively. 
	The insets I1 and I2 of panel (a)~[inset of panel (b)], exhibit(s) the bulk-gap around quasi-energy $E(k)=\pi$ and $E(k)=0$, respectively~[quasi-energy $E(k)=\pi$]. In panels (c) and (d), we illustrate the quasi-energy spectra $E_m$ as a function of the state index $m$ while system obeys OBC, corresponding to panels (a) and (b), respectively; while in the insets of these panels, we depict the 
	zoomed-in quasi-energies for better clarity. Here, we consider  $N=4000$ lattice sites. All the other model parameters are chosen as $(c_{0}, c_{1}, \Delta, B_{x}, t, u) = (1.0, 0.2, 1.0, 1.0, 1.0, 0.5)$.
	}
	\label{Kick}
\end{figure}
After extensive discussions on the generation and characterization of Floquet Majorana modes using periodic step drive protocol, here we introduce periodic kick protocol to generate the Floquet 
$0$- and $\pi$-MEMs in our setup. We consider $H_0(k)$ in between successive kicks, and the on-site kick in the chemical potential is given as
\begin{equation}
	c(t)=- c_{1} \Gamma_{1} \sum_{r=0}^{r=\infty} \delta (t - r T)
	\label{kick_drive_protocol}.
\end{equation}
The Floquet operator in terms of the time-ordered notation can be written as
\begin{eqnarray}
	U_{\rm kick}(k,T) &=& {\rm TO} \exp \left[-i \int_{0}^{T} dt \  \left\lbrace   H_0(k) + c(t) \right\rbrace  \right] \non \ ,\\
	&=& e^{-i H_{0}(k) T} e^{i c_{1}  \Gamma_{1} } \
	\label{kick_evolution_operator}. 
\end{eqnarray}
Having constructed the Floquet operator, we study the bulk quasi-energy spectra of the system. We depict $E(k)$ as a function of $k$ in Fig.~\ref{Kick}~(a) and (b) for driving frequency $\Omega=3.0$ and $\Omega=5.0$, respectively. From the insets I1 and I2 of Fig.~\ref{Kick}~(a), one can notice two bulk-gap openings at the $0$- and $\pi$-gap, indicating two end modes at these gaps. On the other hand, from the inset of Fig.~\ref{Kick}~(b), four gap-openings appear near quasi-energy $\pi$, indicating the presence of four MEMs. Nevertheless, we contemplate OBC to substantiate our predictions, obtained from the bulk spectra. We show the quasi-energy spectra with respect to state index $m$ in Fig.~\ref{Kick}~(c) and (d) for $\Omega=3.0$ and $\Omega=5.0$, respectively. We depict the $0$- and $\pi$-mode eigenvalues in the inset I1 and I2 of Fig.~\ref{Kick}~(c), respectively. Whereas, inset of Fig.~\ref{Kick}~(d) indicates only four $\pi$-MEMs with no $0$-MEM being present in this case.
\begin{figure}[]
	\centering
	\subfigure{\includegraphics[width=0.48\textwidth]{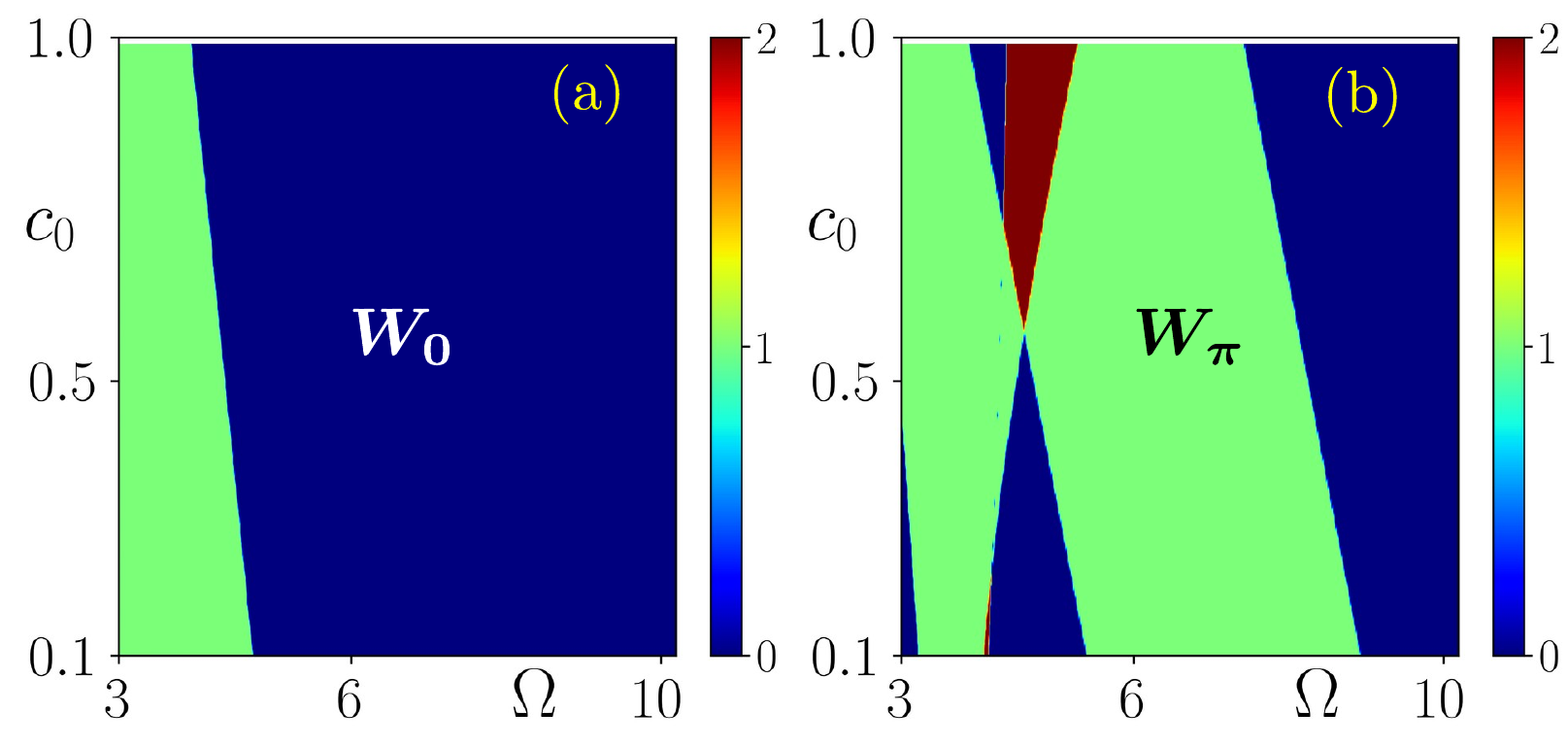}}
	\caption{We demonstrate the dynamical winding number $W_0$ and $W_\pi$ in $c_0 \mhyphen \Omega$ plane in panels (a) and (b), respectively for periodic-kick drive protocol. All the model parameter values remain the same as mentioned in Fig.~\ref{Kick}.
	}
	\label{kickPhase}
\end{figure}

To topologically characterize the Floquet $0$- and $\pi$-MEMs, we calculate the dynamical winding number $W_\epsilon$ defined in Eq.~(\ref{dynamical_wind_step}). We illustrate $W_0$ and $W_\pi$ in the $c_0 \mhyphen \Omega$ plane in Fig.~\ref{kickPhase}~(a) and (b), respectively. Similar to the step-drive, here also, one can generate multiple MEMs. However, this is only possible for the $\pi$-modes (on this parameter range), and the number of multiple modes generated is less than that in the case of step-drive. This can be attributed to the fact that the periodic kick is less efficient to generate longer-range hopping in the dynamical system for this parameter regime, compared to the periodic step-drive protocol. Another point of concern with the dynamical winding number for the periodic kick is that it matches correctly with the OBC quasi-energy spectra only when $\Omega \ge 3.0$. Using the similar line of arguments invoked for the step-drive protocol, we can argue on the Majorana nature of the 
$0$- and $\pi$-end-modes for this case too.

\section{Possible experimental connection}\label{Sec:experiment}
Having discussed the theoretical proposal in engineering the Floquet Majorana modes and examining their robustness under disorder, we here demonstrate the possible experimental connection and its realization. The possible candidate material can be the popular InSb/InAs 1D NW with strong SOC and proximity induced $s$-wave superconductivity (\eg Nb/Al) in it~\cite{das2012zero,Mourik2012Science,DengNano2012,Deng1557,NichelePRL2017,JunSciAdv2017,Zhang2017NatCommun,Gul2018,Grivnin2019,ChenPRL2019,zhang2021large}. 
The time-dependent gate voltage has been theoretically investigated in the context of transport studies \cite{Misiorny18,khosravi2009bound} and can be implemented experimentally~ \cite{dubois2013minimal,Gabelli13} by adopting a suitable superposition of several harmonics. Note that, the first $0<t<T/4$ and last $3T/4<t<T$ parts of the driving scheme include an atomic insulator type model Hamiltonian $H_1$. This can be engineered out of the NW by changing the gate voltage and other control parameters to such a position where bands are relatively flat and substantially gapped. In the intermediate time $T/4<t<3T/4$, the chemical potential can be tuned such that band dispersion in the NW becomes relevant. Experimental implementation of sudden quench between such step Hamiltonians, as demonstrated in Eq.~(\ref{step_drive_protocol}), might be challenging. As a result, there exist a switching time, quantified by the time-gap between two subsequent Hamiltonians, plays crucial role for the practical realization. This switching time scale $\tau$ has to be typically smaller than the than the time scale associated with the Majorana localization length $\lambda$ 
and velocity $v$ of the quasi-particles (Fermi velocity in most of the cases) produced by such quench \ie~$\tau < \lambda/v$. This condition ensures that the MEMs will not diffuse into the bulk during the switching time. The system size $L$ (length of the nanowire) possesses another length scale referring to the fact that $\tau \ll L/v$ as $\lambda/v < L/v$ to avoid overlap of Majoranas inside the NW. 
Note that, once the switching time is greater than the characteristic time scale of the system, the adiabatic evolution will obstruct Majoranas to appear as we have started from a trivial phase. The characteristic time scale is set by the inverse of the gap difference $1/\delta_E=|\Delta E_0- \Delta E_1|^{-1}=f(\Delta, B_x, c_0,c_1)$ associated with the two subsequent step Hamiltonians  $H_0$ and $H_1$.  This condition ensures non-adiabatic evolution in order to conceive the Floquet MEMs. Therefore, we believe that the switching time scale $\tau < {\rm min}\{\delta_E^{-1}, \lambda/v \} < L/v$ in order to observe Floquet topological Majorana modes.

As far as the experimental signature is concerned, implementation of our dynamical protocol might able to signal the topological nature of Floquet MEMs even in the presence of disorder. 
Interestingly, the quantization of the dynamical winding number remains robust for small strength of disorder. Moreover, for intermediate disorder strength, it decreases from the quantized value 
by a small amount which might be possible to probe experimentally (see Fig.~\ref{WindingDisorder}). The $\pi$-MEMs are found to be more robust with disorder in terms of the quantization of the dynamical winding number. Such anomalous mode can only be present in the Floquet topological superconducting phase without any static analog. The driving frequency should be kept at an intermediate value with respect to the other band parameters of the model, such as the hopping and the SOC such that the Floquet topological superconducting phases 
can host multiple number of MEMs within an experimentally conceivable window of the drive parameters (see Fig.~\ref{stepPhase}). For InSb nanowire, the Rashba spin-orbit strength is around $50~\mu{\rm eV}$~\cite{Mourik2012Science}. Following this, our model parameters can take the values: $u \sim 25~\mu{\rm eV}$, $t \sim 50~\mu{\rm eV}$, and $\Delta \sim 50~\mu{\rm eV}$. Thus, the corresponding range of frequency can lie within $\Omega \sim [27 - 76]$ GHz to realize the Floquet MEMs.

\section{Summary and Conclusions}\label{Sec:VI}
To summarize, in this article, we consider a practically realizable model to generate multiple Floquet MEMs based on 1D Rashba NW with proximity induced $s$-wave superconductivity, while an external magnetic field is applied along the $x$-direction. To begin with, we study the topological phase diagram of this static TSC using a winding number exploiting the chiral symmetry. This matches with earlier predictions~\cite{Oreg2010,Leijnse_2012,Alicea_2012}. In this model, a periodic three-step drive protocol is employed to generate the FTSC phase hosting both regular $0$- and anomalous $\pi$-MEMs. Within our driving scheme, multiple MEMs (both regular $0$- and anomalous $\pi$-modes) can be generated. We utilize the periodized evolution operator to contrive the dynamical topological winding number, which can distinctly characterize both $0$- and $\pi$-MEMs. The robustness of the MEMs is investigated in the presence of onsite disorder potential. Nonetheless, the MEMs are almost insensitive to small disorder and remain localized  at the ends of the chain only for specific frequency regime in case of moderate disorder. However, these modes in FTSC phase is destroyed for strong disorder regime. 
In the presence of disorder, the dynamical winding number is computed employing TBC. Given the fact that the topological MEMs are characterized by pfaffian-based $Z$, and $Z_2$ invariant~\cite{XLQi09,XLQi10,Teo10,TewariPRL2012,FZhang13}, here we consider real-space-based dynamical invariant to characterize the Floquet Majorana modes in the presence of disorder. 
We further generalize our scheme of Floquet generation of MEMs for periodic-kick drive protocol. 

We have investigated both the $0$- and $\pi$-MEM wavefunctions corresponding to Fig.~\ref{stepeigen}(c) (step-drive) and Fig.~\ref{Kick}(c) (periodic-kick). We establish the equality condition of the particle and hole part of the wavefunction at a single site. Note that, we can examine the wavefunction only when one MEM wavefunction, localized at one end of the chain,
is separated  from the other MEM state that resides at the other end of the chain. Nevertheless, extracting such wavefunctions by linear superpositions or unitary transformations is not quite 
straightforward for the case of multiple MEMs \ie more than two. The generation of multiple MEMs for a driven system is not unique to first-order TSC~\cite{KunduPRL2013,ThakurathiPRB2013,Molignini18,Yap18} only, but also visible for higher-order TSC~\cite{Ghosh21a,Ghosh21b,Ghosh21c}. On the other hand, static $p$-wave 1D chain 
with long-range hopping involving $n$-th nearest neighbor can host $2n$ number of degenerate MEMs~\cite{DeGottardi13}. Moreover, multiple MEMs are predicted in non-Hermitian systems~\cite{Longwen20}. We can comment that multiple Floquet MEMs can become instrumental for braiding in time domain which is hard to design for static systems~\cite{Bomantara18,BomantaraPRB2018,BelaBauerPRB2019,MatthiesPRL2022}.   

The Floquet generation of MEMs follwing the experimentally realizable NW model has also been investigated in some earlier studies~\cite{JiangPRL2011,YangPRL2021}. In Ref.~\cite{JiangPRL2011}, the authors considered a cold atomic system consisting of 1D NW-$s$-wave superconductor heterostructure in presence of an external laser irradiation to generate Floquet MEMs (both $0$- and $\pi$-modes). Whereas, Ref.~\cite{YangPRL2021} reported the emergence of Floquet MEMs in the Rashba NW model with proximity induced $s$-wave superconductivity. 
The system is made time-dependent via an external ac gate voltage which produces sinusoidally periodic chemical potential. The dissipation and corresponding lifetime of the Floquet MEMs in the 
presence of strong energy and density fluctuations, created by the superconducting proximity effect, have also been investigated. However, in these above studies, the generation of multiple Floquet 
MEMs and their topological protection in presence of random disorder has not been explored. The transport and shot-noise signatures of the $\pi$-modes in Rashba NW setup can also be interesting 
future directions as well.

\subsection*{Acknowledgments}
D.M., A.K.G., and A.S. acknowledge SAMKHYA: High-Performance Computing Facility provided by Institute of Physics, Bhubaneswar, for numerical computations.

\bibliography{bibfile}{}

\end{document}